# Development of a MATLAB/STK TLE Accuracy Assessment Tool, in Support of the NASA Ames Space Traffic Management Project

James MASON

Individual Project Report submitted to the International Space University in partial fulfillment of the requirements of the M.Sc. Degree in **Space Studies**

**August, 2009**

| | |
|---|---|
| Internship Mentor: | **Creon Levitt** |
| Host Institution: | **NASA Ames Research Center** <br> **Moffet Field, CA, USA** |
| ISU Academic Advisor: | **Dr. Hugh Hill** |

# Abstract

In order to improve the effectiveness of conjunction analysis using publically available Two Line Elements (TLEs) a number of strategies are being investigated as part of the Space Traffic Management project at NASA Ames Research Center. To assist in evaluating the effectiveness of these approaches a tool was developed in the MATLAB programming language that interfaces with the AGI Satellite Toolkit and with Microsoft Excel. The TLEs and any available truth ephemerides are read in by the tool and propagated orbits are compared using STK to estimate the errors. This tool is employed to determine the covariance and investigate the growth of errors in propagating the orbit of the CNES Stella geodetic satellite. The different sources of error are assessed and future improvements to the tool are suggested.

# Table of Contents

# List of Figures

# List of Tables

# 1   Introduction

The need for Space Situational Awareness (SSA) has risen in profile in recent months due to a number of high impact events such as the Chinese and US anti-satellite missile tests, the February 2009 collision between the Iridium 33 and Cosmos 2251 satellites and also recent ISS debris alerts.

Earth orbit, and specifically sun synchronous LEO orbit, is a limited natural resource and it is becoming increasingly clear that to avoid orbital debris rending whole orbits unusable there is a need for timely and reliable conjunction analysis on all space objects. Although individual nations, agencies and operators are currently performing conjunction analysis, this is in most cases only for their own assets.

However, it is in the interest of all parties to mitigate the effects of orbital debris. Recent models have shown that debris populations will grow even with no new launches through the generation of the cascading Kessler Effect, where collisions produce new fragments that in turn increase collision rates (Liou, 2007). It is therefore important to leverage debris monitoring efforts with innovative approaches and partnerships. The desirability of maintaining space as a safe operating environment has driven the formation of a number of national debris offices, but efforts are still duplicative and fragmented. To avoid debris becoming a "tragedy of the commons" international transparency and cooperation, specifically on data sharing, would go far to enhance debris models and conjunction analysis.

A key technical requirement is high accuracy debris tracking, which has traditionally been conducted by military radar assets. Although the capability exists, this information is not made freely available and civil and private entities are often left to manage debris risk with unsatisfactory orbital data. The decision to maneuver a satellite is only taken for high risk levels and after repeat observations – moving prematurely may simply put the satellite at greater risk. Often the uncertainty is too great to warrant a maneuver and operators are left to hope for no collision to occur. As a result, collisions have occurred that were not predicted and almost certainly operational capability has been wasted maneuvering spacecraft that, because of poor data, had little real collision risk.

The need for repeat observations (to reduce uncertainty) and the poor performance of propagation models means that the decision to maneuver a satellite is taken at the last minute – at most a couple of days before the collision is predicted. A conjunction that is predicted further in advance would allow less propellant to be used for an equivalent avoidance maneuver.

The primary task of this research is to address and investigate the issue of orbital TLE data accuracy. If a reliable method can be found to reduce uncertainties then it will certainly prove useful in managing debris collision risk. Similarly, if the uncertainty can be modeled it will allow conjunction analysis to be conducted with appropriate error ellipses, instead of "hard" miss distance ellipses or spheres. Secondly, this research will investigate the use of alternative propagator models to see how or where they may be applicable.

This report presents the method and machinery developed to conduct this research, which is deemed the significant product of the internship. This machinery provides a rapid development environment for testing and comparing accuracy assessment and enhancement strategies to identify those with the most promise of delivering real conjunction analysis improvents.

## 1.1 Space Traffic Management Background

## 1.2 Literature Review

An interesting analysis of recent debris collisions can be found at CelesTrak.com, including the current debris clouds from collisions such as the Iridium-Cosmos. Figure 1-1 shows this debris, only 6 months after the conjunction, and it is clear how the debris has spread out to occupy a number of orbital planes, demonstrating how such an event can significantly degrade the orbital operating environment.

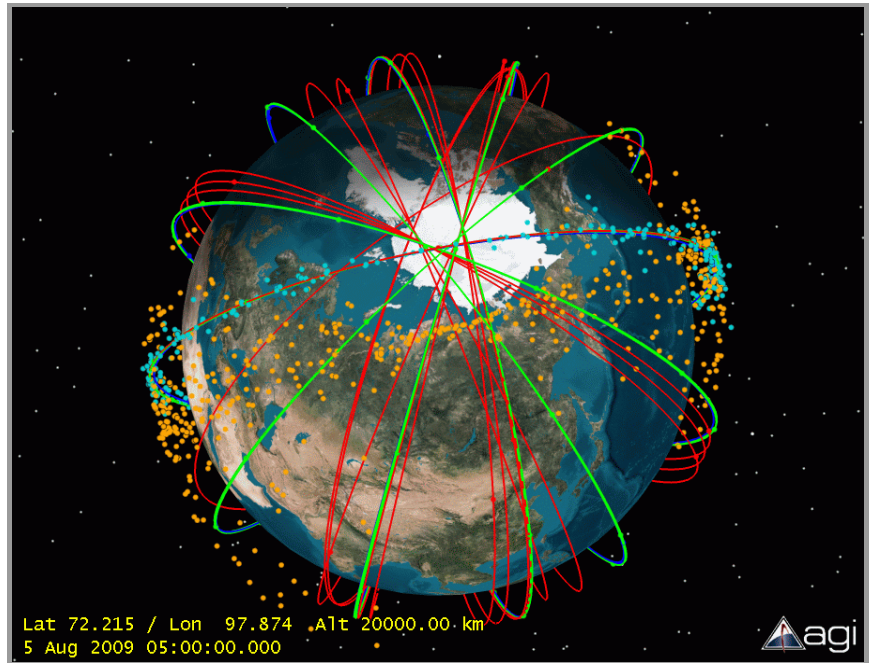

Figure 1-1: Iridium – Cosmos debris cloud. Orange points are debris from Cosmos 2251, blue from Iridium 33. (CelesTrak.com, 10 Aug 2009)

It is often assumed that halting the addition of any new debris objects would stabilize or even reduce the risk associated with debris conjunctions. However, in certain LEO orbits whose satellite number densities exceed a critical spatial density, the creation of new debris through breakups and collisions actually exceeds the loss through orbital decay (Liou, 2007). This LEO window, at around 900-1000km altitude, is currently the region of maximum debris density. Kessler originally proposed the concept of critical spatial densities leading to a 'domino effect' of cascading collisions (Kessler 1991) and this forms the major justification for calls for active debris removal concept studies (Liou, 2008).

The Space Surveillance Network's Two Line Elements (TLEs) are the most comprehensive debris orbital elements source and a number of groups are working on determining the errors of these 'elsets' to improving conjunction analysis through bias corrections and covariance matrix generation.

Kaya and Snow used the GOODOB and MAESTRO programs to screen for 'bad' TLEs and to produce propagation accuracy statistics respectively (Kaya, 1999). GOODOB propagates individual TLEs backwards, comparing them against the previous TLE to identify statistical outliers. MAESTRO, like the method presented here, applies the basic pair-wise differencing approach (see Section 2.1) to generate time-varying standard deviations for a satellite based on historical data. The Aerospace Corporation developed a similar approach in the COVGEN program, which also uses pair-wise differencing to estimate errors (Peterson, 2001). For high altitude MEO and GEO objects COVGEN uses operator data to estimate initial errors. By similarly assuming a normal distribution of errors, COVGEN determines 'reasonable estimates' for about two thirds of the space catalog.

The Satellite Orbital Conjunction Reports Assessing Threatening Encounters in Space (SOCRATES) system also uses pair-wise differencing to estimate the covariance of a TLE. It additionally makes a simple correction for bias before using the resulting 'corrected' standard deviations to propagate satellites' orbits (with related errors) for conjunction analysis (Kelso, 2005). The major assumption, in comparing TLEs to TLEs with pair-wise differencing is that there is minimal prediction error at the epoch of the TLE – assuming that TLEs can be used as a 'truth' is not necessarily valid and will be addressed later in this report. Kelso has gone on to compare pair-wise differenced TLEs with GPS ephemerides for 22 GPS satellites (Kelso, 2007). While this somewhat validates this approach, and suggests improvements for the COVGEN approach, the MEO GPS orbital regime differs greatly from the LEO maximum debris risk region and generalizations across regimes may not be valid.

The ESA/ESOC Space Debris Office has also generated covariance information using TLEs with the ODIN program (Flohrer, 2008). ODIN's accuracy for orbit determination and propagation was validated by post-processed orbital data for Envisat. Flohrer et al. note in their paper that the optimal approach is to compare covariance information against operator or post processed truth data, but the number of satellites for which this is possible is limited.

## 1.3 Space Traffic Management at NASA Ames

### 1.3.1 STM Project History

NASA Ames initiated a small research effort in the area of space traffic management in 2006 which was primarily focused on providing a broad overview of the technical possibilities and challenges of traffic rules for the space environment and applying the lessons learnt from Ames' expertise built up in developing future Air Traffic Management systems (for the FAA).

Following this initial work, Ames sponsored a study project on STM at the International Space University in Beijing in the summer of 2007. The project involved a team of approximately 30 students from various backgrounds whose aim was to develop a first technical design for an STM system for Earth orbit. This project produced a first conceptual design for an STM system, with particular focus on conducting all-on-all conjunction analysis and collision avoidance, and suggesting a system of orbital slots for SSO. In addition, the report looked at the international institutional mechanisms for implementing such a system and the related costs.

Following the ISU report NASA Ames initiated a research project at the University Affiliated Research Center (UARC). UARC funded two projects, one on mathematical techniques to correct the NORAD TLEs and the other on an economic (game-theoretic) analysis of systems to incentivize the mitigation and removal of space debris. This work was carried out by two faculty and two Masters students from UC Santa Cruz Department of Engineering.

In 2009, Ames embarked upon a more concerted effort in the area of satellite collision avoidance following the Iridium-Cosmos collision. The goal was to see how and whether NASA can help to ensure that such collisions do not occur in the future, where they are avoidable. The research effort, encouraged by USAF personnel, is aimed at two key technical challenges: setting up and running all-on-all object conjunction analysis on the NASA Advanced Supercomputing (NAS)

Division's Pleiades supercomputer (and the display of the results on the Ames Hyperwall) and the improvement of TLE prediction accuracy.

### 1.3.2 STM Project Progress Report

The NAS supercomputer is currently set up to do a number of interesting simulations. These include all-on-all conjunction analysis of the whole space catalog, with each conjunction being displayed in real time on each of 128 monitors comprising the hyperwall. This has additionally been done for a theoretical future catalog with over 3 million objects down to 2 cm size. This demonstrates that Ames has the computing power to do all-on-all conjunction analysis if the right data is available. Since the military will not currently allow other entities access to their high accuracy catalog users are forced to work with the low accuracy TLEs, hence the need for analysis of TLE accuracy and formulation of corrections strategies.

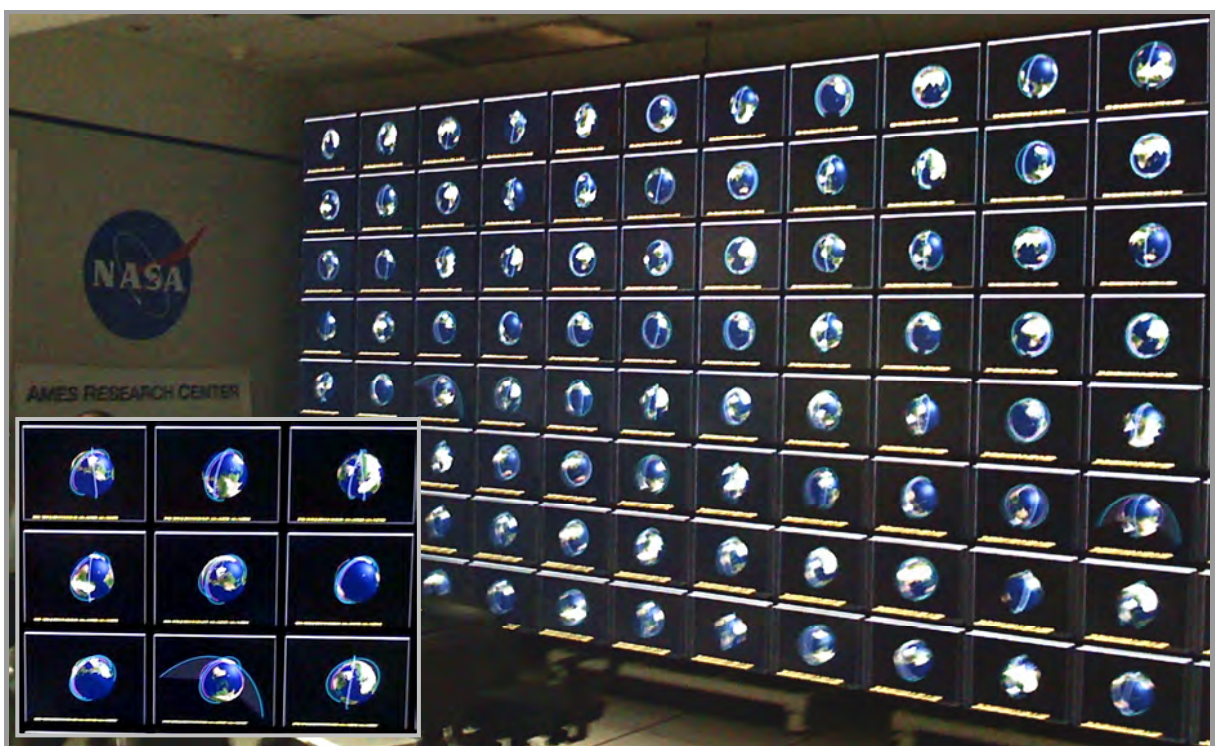

Figure 1-2: The NASA Ames Hyperwall runs off the Pleiades supercomputer – with multiple coordinated monitor displays (each with its own GPU) it has over 64 million pixels and is capable of displaying 100 Gb/s for interactive exploration of multidimensional or multivariate.

Efforts are underway at Ames and at UC Santa Cruz to predict future TLEs by fitting curves to historical data. These average out any short term perturbations and appear more promising than the standard – Simplified General Perturbations (SGP4) - model at predicting TLEs out to a few weeks. Alana Muldoon, of UC Santa Cruz, presented her ongoing work (along with the author) at an STM workshop hosted by Ames on 30 July 2009 (Muldoon, 2009). This somewhat addresses divergent propagator models, but does not address the inherent uncertainties in the TLEs. Ames and UC Santa Cruz are also working on determining TLE corrections in relation to truth orbits. Muldoon and Elkaim at UC Santa Cruz have determined a global rotational bias between the TLEs and the GPS truth data in three of the six GPS inclination bands, but this remains unverified at this moment (Muldoon, 2008). Similarly, there are indications that fitting simple sinusoidal curves to a TLE propagated with the SGP4 model (see Section 1.4.1) could provide a correction that lowers the variance from truth by a factor of 3.

### 1.3.3 The Role of this Analysis in the STM Project

The research topic presented here developed out of discussions with a numbr of Ames staff. A review of the high level goals of the project and the current weaknesses led to the identification of three key questions:

- What is the accuracy of the TLE sets themselves?
- How can we improve the accuracy of the TLEs?
- How can we improve the propagation accuracy of TLEs?

As a result of these questions the required work was segmented into four tasks, defined in Table 1-1. Each task was designed to provide some further understanding of the technical aspects of the three key questions and all were implemented in a single program using MATLAB and the AGI Satellite ToolKit (see Section 2.6 for implementation details).

Table 1-1: Proposed task list

| # | Prop. Model | From Epoch | To Epoch | Methodology | Objective | Comparison |
|---|---|---|---|---|---|---|
| 0 | SGP4 | TLE | TLE | Pair-wise Differencing | Covariance of SGP4/TLE | SGP4/TLE vs. TLE |
| 1 | SGP4 HPOP | TLE POE | TLE TLE | Pair-wise Differencing Epoch Synching | Covariance of SGP4/TLE around truth | SGP4/TLE vs. POE |
| 2 | HPOP | TLE (state) | TLE | Pair-wise Differencing | HPOP as a TLE propagator | HPOP/TLE vs. TLE |
| 3 | SGP4 | TLE | POE | Single Propagation | Short term perturbations in SGP4 | SGP4 vs. Truth |

**Task 0**
Task 0 is named as such because it is essentially a repetition of a method that has been demonstrated elsewhere a number of times. With pair-wise differencing, each TLE is propagated using the simplified general perturbations (SGP4) model to the epoch of all future TLEs in a time period and residuals are computed in the satellite coordinate frame (see Section 1.4.1 for more details on TLEs and the SGP4 model). By determining the variation of the TLEs around each other, the errors (at epoch) in the TLEs can be estimated and covariance matrices generated. This will allow an error ellipsoid to be propagated with the satellite position. The benefit to repeating this method is twofold: Firstly, it provides Ames the technical capability to reproduce the efforts of other groups locally. Secondly, it sets a baseline method to be adapted in the future tasks and to use as a comparison. The exact pair-wise differencing methodology is explained in Section 2.1.

**Task 1**
A variation on the above, Task 1 repeats the pair-wise differencing method but compares the SGP4 propagated TLEs with post processed truth data. The advantage of this is that the estimated errors and the covariance matrices represent variance around where a satellite actually is, not just where TLEs report the satellite to be. This is more relevant than the previous task, since the goal is to avoid real conjunctions. Like Task 0, this method can also detect biases in the data and a simple bias correction could be applied to reduce the overall variation. Although this can only be done for satellites where truth data is available, it is important to set up this machinery for future work. As the correction work at Ames and UC Santa Cruz matures, methods of refining or correcting TLEs will need verification. These corrected TLEs can be fed into the algorithm and a quantitative analysis done of the improvement in covariance as compared to truth.

**Task 2**

Task 2 is primarily a demonstration of the inclusion of an alternative propagator in the in the MATLAB program. This is done by repeating Task 0, but using STK's High Precision Orbit Propagator (HPOP) to do pair-wise differencing by propagating each TLE position-velocity state vector ($r_x$, $r_y$, $r_z$, $v_x$, $v_y$, $v_s$.) to each subsequent TLE epoch and calculating residuals. While HPOP is not designed for use with TLEs it is in fact highly accurate when initial conditions and parameters are known and may prove useful if initial states and parameters can be estimated (Vallado, 2005). This is theoretically possible by statistical smoothing of TLEs and by backwards parameter optimization techniques and this could be an area of future use for this method.

**Task 3**

Task 3 was to produce high resolution ephemeris data from the SGP4 propagator for comparison to truth ephemerides. This is useful for investigating the nature of short term periodic perturbations neglected in the SGP4 model. This task was effectively handed over to another member of the team due to time limitations.

The amalgamation of these tasks into a single program was designed to create an environment for testing and comparing the various approaches to TLE accuracy assessment and prediction. The final product was to be a MATLAB script that takes in two alternative data sources (TLEs and 'truth' ephemerides), uses two alternative propagators (SGP4 and HPOP) and produces results that can lead to meaningful valuations of the tested approach. However, to demonstrsate this tool's capabilities it is desireable to find an object that has both publically available TLE data and some form of truth data and use the tool to compute and compare errors.

## 1.4 Data Sources

Two data sources were utilized in this study. The North American Aerospace Defense Command (NORAD) developed the TLE format (for format details see Appendix A) as a compressed form of mean classical orbital elements. These are publically available and exist for most space debris down to 10 cm size. The second source is post-processed Precision Orbit Ephemerides (POE) from the Center for Space Research (CSR) at the University of Texas at Austin. These are only available in this form for six satellites.

### 1.4.1 Two Line Element (TLE) Sets

TLEs are a special form of mean classical orbital elements that are expressed in the true equator, mean equinox (TEME) frame. TLEs are generated with an orbit detemrnination process based on observations by the United States Space Surveillance Network (SSN), which comprises a number of radar and electro-optical sensors. Once tracked by the SSN an orbit is estimated using the SGP4 model and expressed in the form of a TLE (Vallado, 2006). Since November 2003 TLEs have been made publically available in the Space Object Catalog by The Joint Space Operations Center (JSpOC). The publically available TLEs include 14968 space objects of 10cm or larger, of which 11524 are debris or rocket bodies (Space-Track Satellite Situation Report, 06 Aug 2009). TLEs are a form of mean classical orbital elements that average over certain short term perturbations. Some short term perturbations are reconstructed when a TLE is propagated using the appropriate (SGP4) model. It has been shown that the SGP4 model implemented in STK closely matches that used in generating the TLEs (Kelso, 2005).

The SGP4 model is an analytical general perturbations model developed by NORAD and NASA. It is applicable for low earth orbit satellites, with periods less than 225 minutes, and works by solving Kepler's equation before introducing specific long and short-period periodic perturbation terms.

TLE data can be freely downloaded from the Space-Track.org or CelesTrak.com websites. TLEs are currently the most comprehensive and useful publically available catalog of orbital debris orbits. However, JSpOC does not publish their estimated accuracy for these TLEs and it is clear that TLEs are only useful in predicting orbits for a few days with moderate accuracy (see Section 4). This analysis used TLEs downloaded from Space-Track.org.

### 1.4.2 Precision Orbit Ephemerides (Truth Data)

POE 'truth' data was obtained through John C. Reis at the CSR at the University of Texas at Austin. These were produced in the J2000 coordinate frame as time-stamped position velocity state vectors $(t, \vec{r}, \vec{v})$.

The POEs are produced using various data sources, such as the International Laser Ranging Service (ILRS), the Doppler Orbitography and Radiopositioning Integrated by Satellite (DORIS), on-board accelerometers and, in the case of TOPEX/Poseidon, GPS. These data are post processed at CSR with the product being state vectors accurate down to a few centimeters (Reis, 2009) – well within the limits required for this analysis. Upon request, POE data was provided for the 6 satellites in Table 1-2 with the following properties:

- o Starting Epoch 29 Dec 2003 00:00:00.0
- o Ending time 28 Jan 2005 00:00:00.0
- o 60 second time steps
- o 570,240 ephemeris points

A MATLAB algorithm (ExtractUTPOE.m) was developed to reformat this data as an external ephemeris (.e) file for use in AGI's STK software. This frame was selected since it is clearly defined and avoided any errors in misinterpreting the frame. In addition, the algorithm constrained the data to 126,000 points (87.5 days) after the starting epoch, which was more than sufficient for this analysis.

### 1.4.3 Available Ephemerides and Satellite Selection

CSR maintains ephemerides for six active satellites in the LEO and MEO regimes (see Table 1-2). Since this analysis is in support of space debris conjunction analysis, the most valuable satellite is that whose orbital regime resembles that of common debris objects.

Table 1-2: Properties of satellites in CSR ephemeris database

| Name | Catalog ID | Perigee (km) | Inclination (º) | Eccentricity | Mass (kg) |
|---|---|---|---|---|---|
| Starlette | 07646 | 812 | 49.8 | 0.021 | 47 |
| Lageos-1 | 08820 | 5860 | 109.8 | 0.005 | 406 |
| Ajisai | 16908 | 1490 | 50.0 | 0.001 | 685 |
| Topex/Poseidon | 22076 | 1336 | 66.0 | 0.001 | 2,402 |
| Lageos-2 | 22195 | 5620 | 52.6 | 0.014 | 405 |
| Stella | 22824 | 800 | 98.6 | 0.021 | 48 |

Of particular interest are Stella and TOPEX. Stella resides in a sun synchronous LEO orbit and the highest debris spatial density is found in this orbital regime at about 900-1000km altitude. Stella, launched in 1993 by CNES, is a low area-to-mass (low drag) passive geodetic spherical satellite with a diameter of 24 cm. It is covered in 60 retroreflectors to allow accurate laser ranging which means that very good ephemeris data is available. The orbit and size of Stella make it the closest substitute for an actual debris object and therefore good for an analysis of this nature.

The Ocean TOPography Experiment (TOPEX) is 2 ton satellite with dimensions 5.5m by 2.8m. Its larger mass and size might more closely simulates larger orbital debris, as well as active or retired mission payloads. TOPEX is fitted with a retroreflector array, allowing precise ephemerides to be generated.

Stella has been selected as a case study for this report, to provide a debris like object to demonstrate the accuracy assessment tool developed here. This tool could be applied to any object for which sufficient data exists however.

# 2 Methodology

The methodology used in this analysis is adapted from work by Osweiler in his Masters thesis in 2006 and a paper by T.S. Kelso (Kelso, 2005). The method allows for the variance of a set of TLEs to be investigated for a single object as well as for the formation of covariance matrices. The validity of this approach, specifically the applicability of Estimation Theory relating to this method, focusing on the Central Limit Theorem and the Principle of Maximum Likelihood, can be found in Osweiler's thesis.

In this analysis the method can be applied to compare and quantify the difference between propagation models. Similarly it can be used to quantify improvements in accuracy provided by some kind of corrections or smoothing of TLE data. Finally, it can be used to develop both covariance matrices and time varying statistics which may be used to develop error ellipsoids for use in conjunction analysis. These outcomes are produced through fulfillment of the four tasks discussed in Section 1.3.3.

## 2.1 Basic Pair-wise Differencing of TLEs

Given a number of TLEs over some time period for one object, we expect them to have some error or variation around where the object really is, due to tracking and orbit determination errors. The variation in the TLEs can be compared only at the same time, so it is necessary to propagate the TLEs to a common comparison epoch. With a perfect satellite propagator, the variation in position between the series of TLEs propagated to a common reference time reveals the variation and distribution of the TLEs themselves. However, any propagator represents a simplified model of reality and therefore introduces further variation, depending on initial conditions and parameters, and this effect will increase with the time propagated. Propagating a series of TLEs to a reference time therefore provides an estimation of this variance and similarly a covariance matrix can be generated. The covariance matrix is the generalization to higher dimenstions of the scalar variance – the diagonals give the scalar variances and the off-diagonals give the covariance between different vector elements.

The pair-wise differencing method, as demonstrated by Osweiler, effectively provides a way to determine variance of TLEs around themselves but reveals nothing about systematic biases between the published TLEs and where the satellite actually is. However, it uses only the data that is publicly available for orbital debris objects (TLEs) and can therefore be applied across the whole catalog.

The basic approach is, for some given time period, to loop through each TLE in the period and to set each TLE as an initial condition in a propagator model. As mentioned previously, the appropriate model for use with TLEs is the SGP4 model. STK/SGP4 is used to propagate the TLE until the end of the time period. Once this has been done for all TLEs in the period, a second loop extracts the state vectors at each TLE epoch from all of the propagated orbits to calculate residuals. In basic pair-wise differencing the residual is the difference between the propagated state vector $S_j$ and the state vector $S_k$, where $S_j$ has been propagated for time $\Delta t = t_k - t_j$. At time $t_k$ the best estimate of the actual state is $S_k$, so the residual can be expressed more generally as

$$\partial S_j = S_j - \bar{S} \qquad (2\text{-}1)$$

where $\bar{S}$ is the assumed 'best estimate', in this case $\bar{S} = S_k$, the un-propagated TLE.

Assuming a time period *T*, which has *N* TLEs numbered from 1 to *N*, the simplified algorithm looks as follows:

```
        For TLEs in time period T
 ┌─ Loop backwards through TLEs, from j = N to j = 1
 │       Propagate TLE(j) to EndTime with SGP4
 │    ┌─ Loop backwards through TLEs, from k = N to k = 1
 │    │      If (k>j) then:
 │    │      Determine position of TLE(j) at epoch of TLE(k)
 │    │      Calculate residuals between TLE(j) and TLE(k)
 │    └─ End loop
 └─ End loop
        Determine Statistics
        Display Results
```

The comparison of TLE(j) propagated to the time of TLE(k) yields a position residual, in this case calculated in the Radial, Transverse and Cross-track (RTC) reference frame, which is often used to describe orbital errors and relative positions (see Appendix B for details on this frame).

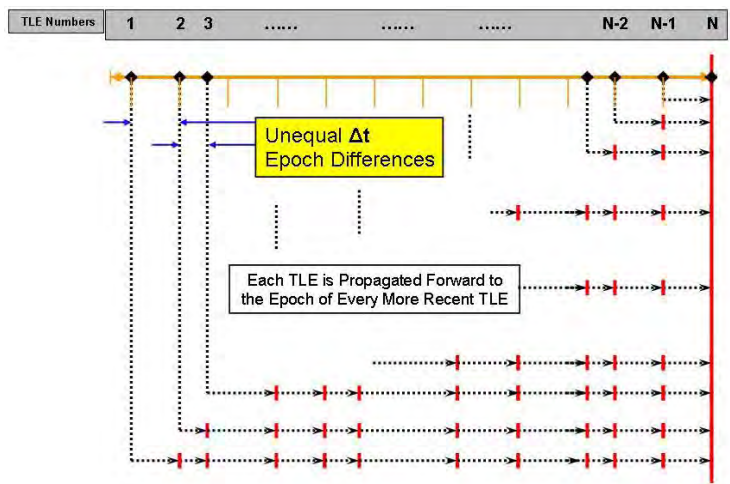

Figure 2-1: Pair-wise Differencing Method (Osweiler, 2006)

The pair-wise differencing method is represented diagrammatically in Figure 2-1. Since TLEs are not produced at regular time intervals, each residual is associated with some propagation time, Δt. This allows residuals to be binned according to propagation time so that some statistical analysis can be performed, including calculating the mean, variance and standard deviation of each bin and the probability density distribution (if a normal distribution is assumed).

## 2.2 Pair-wise Differencing of TLEs with Truth

As a variation on the basic pair-wise differencing scheme, one may wish to compare the TLE data with post processed truth data in the form of POEs. This allows investigation of the error or variance of TLE data around a satellite's actual position, instead of around the position reported in the TLEs. The method is very similar to that described above, with the addition of another propagation routine. Since pair-wise differencing computes residuals at the epoch times of the TLE data we need to ensure that the truth POE data is synchronized with the TLE epoch times. Since the step size in the POE data is 60 seconds, the maximum propagation required (for a POE data point immediately prior to the TLE epoch time) is 60 seconds. An early analysis showed for 60 second propagations, an HPOP orbit remains less than 60cm from SGP4. We can therefore assume the synchronized ephemeris to be a maximum of 60cm from the real satellite position, well within acceptable limits for this analysis.

```
            For TLEs in time period T
    ┌─ Loop backwards through TLEs, from j = N to j = 1
    │           Find nearest POE and propagate to epoch of TLE(j)
    │       ┌─ Propagate TLE(j) to EndTime with SGP4
    │       │  Loop backwards through TLEs, from k = N to k = 1
    │       │          If (k>j) then:
    │       │          Determine position of TLE(j) at epoch of TLE(k)
    │       └─         Calculate residuals between TLE(j) and POE(k)
    │           End loop
    └─ End loop
       Determine Statistics
       Display Results
```

The residuals between TLEs and truth are calculated as previously, except that in this scenario $\overline{S}$ is given by the nearest previous POE propagated to time $t_j$. These residuals are more enlightening than the residuals between TLEs and TLEs – they give the actual variation of the TLEs around the satellite position. We can expect some measurement error in the TLEs and this will lead to a more dilute distribution. Global correction schemes and TLE 'smoothing' may lead to a reduction in this error and therefore produce narrower distributions, which is very beneficial for conjunction analysis. This method can therefore be used to test corrected TLEs to determine if, and by how much, the variance has been improved by the correction. The weakness in this method is the reliance on truth ephemerides, which are not widely available for active satellites and not at all available for space debris.

The case study of the Stella satellite will begin with this comparison of propagated TLEs against truth. Since truth is known for Stella, it is natural to determine the error around the known truth for this satellite and then generalize by doing the analysis without truth data for comparison. Further analysis is necessary before any generalizations can be made across orbital regimes about the variance of TLEs around the truth, but Stella serves as a demonstrator of the general method.

## 2.3  Pair-wise Differencing of TLEs using HPOP

The methodology employed here is identical to the basic pair-wise differencing except that the TLEs are propagated using STK's HPOP propagator. As mentioned previously, TLEs are designed for use with the SGP4 model, which reintroduces missing short-term perturbations into the orbit. However, it is important firstly to demonstrate that other propagators can be evaluated using this tool and secondly that other propagators are not appropriate for propagation of TLEs. This method will reveal the extent of HPOP's inadequacy for propagating TLEs while also providing the algorithm for possible future work.

```
            For TLEs in time period T
    ┌─ Loop backwards through TLEs, from j = N to j = 1
    │           Propagate TLE(j) to EndTime with SGP4
    │       ┌─ Loop backwards through TLEs, from k = N to k = 1
    │       │          If (k>j) then:
    │       │          Determine position of TLE(j) at epoch of TLE(k)
    │       │          Calculate residuals between TLE(j) and TLE(k)
    │       └─ End loop
    └─ End loop
       Determine Statistics
       Display Results
```

To propagate TLEs with HPOP the TLE is converted in to a position-velocity state vector at the time of the TLE epoch. The state is then propagated using HPOP and residuals generated against the other TLEs (that have been converted to states). HPOP is highly sensitive to the parameters used and this means that to accurately propagate a satellite some satellite specific parameters need to be known, such as area to mass ratio and drag coefficients (Vallado, 2005). Even more significant are the initial conditions used with HPOP. The inherent uncertainty in TLEs means that the initial conditions are highly variable and this may cause rapid divergence when propagating. However, in the future, if a high accuracy catalog becomes available or if a smoothing/correcting scheme can provide accurate initial conditions then the HPOP method may prove useful as a "TLE propagator".

## 2.4 Statistical Analysis

The same statistical analysis was performed for the discussed methods. This was to divide the computed residuals into bins according to propagation time before computing the mean, variance and standard deviation for each bin. Additionally the covariance matrix was calculated using the same method as that employed in Osweiler, 2006.

It is apparent in Osweiler's work that calculating residuals from the beginning to the end of a period leads to a trailing off in the data. This is a direct result of the fact that the first TLE, propagated to the end time, provides only a single point with propagation time $\Delta t_{max}$ whereas in the middle of the data we have many TLEs that are propagated for short periods of time (see Figure 2-1). In order to generate more points and to attempt to avoid this weighting of bins Osweiler uses a loop of multiple time periods to calculate residuals.

Due to the difference in implementation, the algorithm presented here computed Stella propagations for a period of 30 days, but constrained residuals to propagation times of up to 7 days. This is effectively a moving window of 7 days that results in more residuals being generated and a more even weighting throughout the bins. Both of these periods can easily be extended to generate more points and for longer propagations, but this was deemed sufficient for this analysis.

The choice of bin size affects the resulting statistics and this bias-variance tradeoff is a central dilemma of density estimation (an introductory treatment of this can be found on the AI Access website). Larger bins provide better bin statistics but a poorer picture of the distribution (Osweiler, 2006). The bin size for this Stella analysis was selected empirically to provide a statistically valid number of points per bin, while maintaining an acceptable resolution across the time range. A bin size of 0.58 days (12 bins) resulted in bins containing 31 data points each on average.

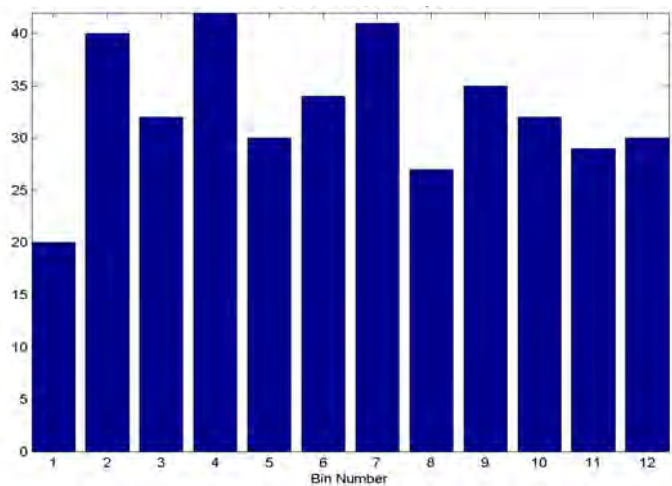

Figure 2-2: Histogram showing the number of points in each bin
for the Stella geodetic spacecraft, with a bin size of 0.58 days.

## 2.5 Covariance Matrix

The covariance matrix of an object can be thought of as the generalization of variance to higher dimensions and it is used to compute the object's positional probability distribution. The covariance of state $S$ is defined, in matrix notation, by

$$\Sigma = E[(S-m)(S-m)^T] \qquad (2\text{-}2)$$

where $m = E[S]$, the expectation value of $S$. Alternatively, $m$ can be thought of as 'a vector comprised of the mean elements of the respective elements of the residual state vectors' (Osweiler, 2006). The scalar form of $m$ is simply the scalar mean $\mu = E[S]$, and the scalar standard deviation and variation is given by

$$\sigma^2 = \text{var}(S) = E[(S-\mu)^2] \qquad (2\text{-}3)$$

For a state vector $S = (r_x, r_y, r_z, v_x, v_y, v_z)$, the covariance will be a 6 x 6 matrix with the matrix diagonal giving the variances of each state element while the off-diagonal elements give the cross-correlation between elements. The covariance of an object can be propagated from time $t_0$ to time $t$ by using the state transition matrix $\Phi(t,t_0)$, which is determined by solving the differential equations of motion for a given force model. It therefore depends on the propagator selected. Error growth is then a combination of initial state vector covariance and errors growing because of an inaccurate force model in the propagator.

Finally, the growth of errors can be investigated by centering the mean, variance and standard deviation at the middle of each bin, allowing a least squares analysis to determine error and bias growth behavior.

## 2.6 Implementation

This analysis was performed using a personal Toshiba Satellite-Pro laptop and using the following software tools:
- Analytical Graphics Inc. Satellite Toolkit (STK) 8.1.1
- MATLAB 7.5.0
- MS Office Suite 2003

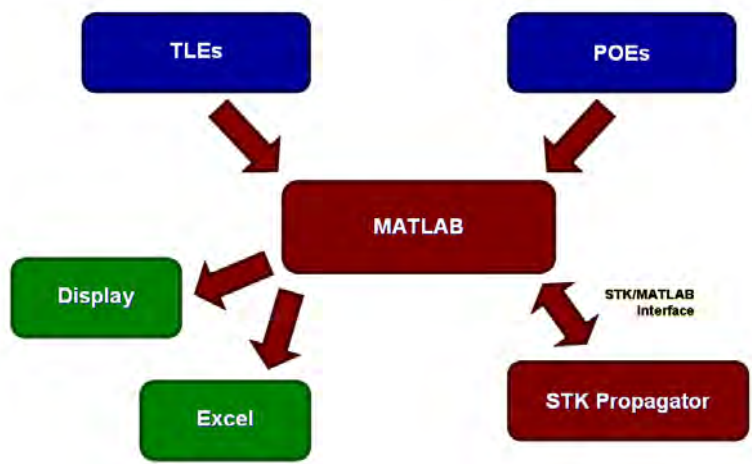

Figure 2-3: MATLAB/STK Tool Elements. TLEs and POEs files are read in by MATLAB and the interface used to connect to the STK propagators. Calculations are done with MATLAB and results displayed as figures or saved to Excel data files.

This analysis was conducted through a MATLAB script that reads in and converts the two data forms and utilizes the MATLAB/STK interface to set up satellite propagations, perform coordinate transformations and output results (see Figure 2-3). MATLAB can display the results to the screen, or save them in the form of Excel data files or figures.

### 2.6.1 STK/MATLAB Interface

The STK/MATLAB interface is initialized with the command agiInit, which enables the mexConnect toolset. mexConnect is used to establish a connection with STK to set up STK scenarios and send simple commands to STK such as basic propagations. Additionally, MATLAB's aerotoolbox allows for a number of aerospace operations, such as basic propagations, coordinate transformations and TLE handling.

However, to expose the full functionality of STK it was necessary to use the stkConnect command. stkConnect allows almost all of the functionality of STK to be called from the MATLAB interface. This includes more advanced propagation (such as setting HPOP force model parameters), controlling graphic settings and exporting reports and ephemeris files. In order to maintain consistency in propagation models, stkConnect was used as a standard command for calling STK functions even when it was found to be less efficient (for more details see 'Connect Commands' in the STK Help)

### 2.6.2 Other Functional Elements

In addition to the bespoke MATLAB script that was developed a number of other functions were developed or acquired. As part of this analysis the function POEtoSTKe was developed that reformats CSR provided POE data into STK's external ephemeris file format by rearranging the data and adding the appropriate headers. This is useful for any STK analysis using the CSR data.

The 'fastsearch' algorithm by Valentin Kuklin was used to rapidly search through large datasets to identify the POE preceding each TLE time. A number of other functions that are available for download as companion code to the book 'Fundamentals of Astrodynamics and Applications' were called by the main script (Vallado, 2001). The most important of these were rv2rtc, which convert from position-velocity state vectors to RTC frame and covct2cl which converts a covariance matrix from Cartesian XYZ to classical orbital elements.

### 2.6.3 Algorithm PsuedoCode

Before MATLAB coding commenced a pseudo code was developed to check the logic used in formulating the algorithm and to use as a blueprint while developing the main body of the program. This pseudo code, shown below, was then implemented in the MATLAB script MASON09.m (see Appendix C). The numbers in parenthesis on the right show the relevant tasks for a specific piece of code (see Table 1-1).

```
    Initialize simulation options: start/stop times, bin sizes, etc
    Initialize MATLAB variables
    Connect STK/MATLAB interface and create scenario
    Reads TLEs from file if they fall in simulation time period
    Read in CSR POE file and convert to STK ephemeris file
    Create "truth" satellite from created STK ephemeris file

┌ Loop backwards through TLEs, from j = N to j = 1
│       Connect to STK and create satellite j
│       Propagate TLE(j) to EndTime with SGP4                    (0,1,3)
│       Convert TLE(j) to state vector S_j                       (2)
│       Propagate state vector S_j to EndTime with HPOP (2)
│       Propagate preceding POE to TLE(j) to create Truth(j)     (1)
│       If j=1: export small step size ephemerides for TLE(1)    (3)
│
│     ┌ Loop backwards through TLEs, from k = N to k = 1
│     │     Check to see if sat exists already (k > j)
│     │     Calculate residuals for TLE(j) vs TLE(k) with SGP    (0)
│     │     Calculate residuals for TLE(j) vs TLE(k) with HPOP   (2)
│     │     Calculate residuals for TLE(j) vs Truth(k)           (1)
│     │     Separate data according to binsize
│     │     Create large data array for export
│     └ End loop
└ End loop

┌ For Task 0, 1 and 2
│     Calculate mean, standard deviation and variance of each bin
│     Fit polynomials to mean, std dev and variance
│     Calculate covariance matrix in satellite frame
│     Convert covariance to classical elements
│     Save data to Excel file
│     Display various plots/figures
└ End
```

### 2.6.4 Computational Time

Running a 30 day simulation, with all options activated, on a personal laptop for the Stella satellite took 98 minutes. This is unacceptable for implementing across the whole catalog of about 15,000 objects, but the algorithm used is not optimized for fast computation. The current implementation was designed to create a data exploration environment, where many unrelated quantities are calculated. When a particular strategy is shown to reduce uncertainty and can be generalized for all objects then the algorithm can be trimmed and optimized for application on the supercomputer and for the all-on-all problem.

# 3 Propagated TLEs against Known Truth Data

## 3.1 Introduction to Task 1

For conjunction analysis it is important to the operator to know the errors associated with a satellite's state that has been generated with a chosen model. The obvious way to determine this is to compare the model with actual truth data. This is the initial approach taken in the selection of Stella for this analysis.

## 3.2 Assessing the Accuracy of the TLEs

The TLEs are an approximate measurement of the actual position of the satellite. Of primary interest is the uncertainty of these TLEs around the actual position. By converting the TLE to position velocity state vectors they can be compared against the POE truth ephemerides (that have be synchronized to the TLE epoch time). The range differences between TLEs and truth for Stella in Jan 2004 are shown in Figure 3-1 – for the period of this analysis the TLEs are on average 600m away from truth, although some are as much as 1.4 km from the actual satellite position. The standard deviation for this satellite is about 340m.

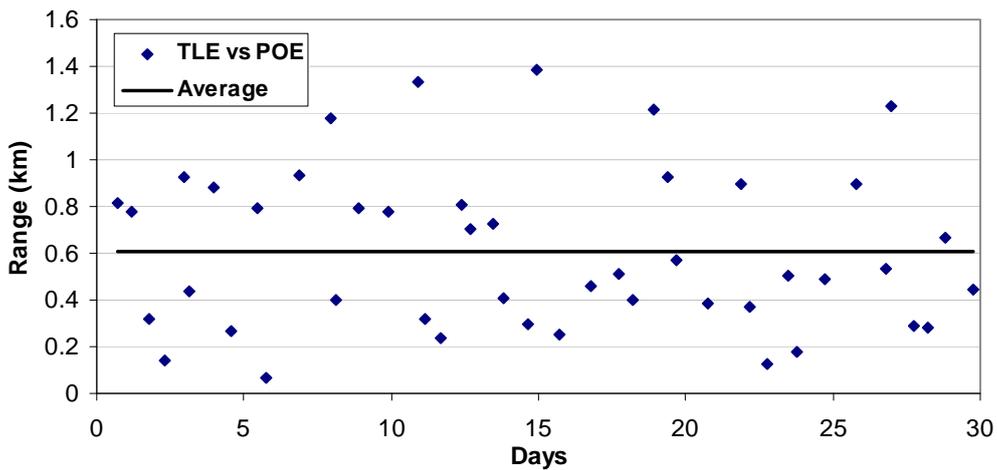

Figure 3-1: Range residuals between published TLEs and POEs synchronized to the TLE epoch

This uncertainty in the actual TLEs is one distinct source of error that arises from two separate effects. Firstly, the observations by the Space Surveillance Network of the satellite's position are not exact. Secondly, the orbit determination model used to generate the TLEs themselves is not an exact model and these both introduce the error in TLEs compared against truth. The second source of error that is of interest is that introduced by an inexact propagator model. When a satellite's future position is predicted there is some uncertainty that is expected to grow according to the time ahead that is being predicted, simply because the model used for prediction is not perfect. For conjunction analysis, in predicting future satellite positions, it is difficult to separate out these error sources. However, these errors are not independent – the bigger the error in initial condition, the bigger the propagated error is expected to be. By pair-wise differencing these TLEs and comparing them to truth the accuracy of the TLE/propagator regime can be assessed.

## 3.3 Results

These results are for Stella for the period of Jan 2004, where propagated TLEs are compared against POEs.

### 3.3.1 Residuals

It is expected that an SGP4 propagated position will diverge from the updated TLEs and this is confirmed in SOCRATES (Kelso, 2005). The residuals in RTC between propagated TLEs and truth are displayed in Figure 3-2. The first source of error – the variance in the actual reported TLEs – is apparent for very short propagation times (where we would expect very little error). In the radial direction, there appears to be a definite secular bias in the data. However, this is not apparent in the along-track and cross-track directions. Detecting, modeling and removing this bias is one way of improving the accuracy of TLE propagation by predicting how the error grows.

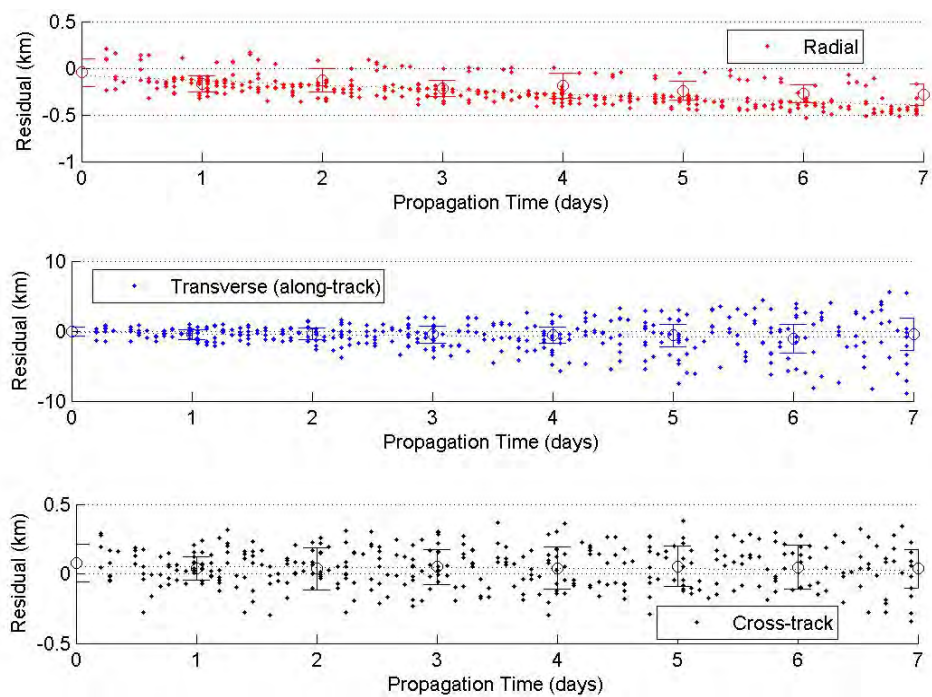

Figure 3-2: Stella RTC position residuals between propagated TLEs and truth for 7 days of propagation

The error arising from the propagator is evident as the propagation time increases and the variance, in general, grows. However, the error appears to stay roughly constant over 7 days of propagation in the radial direction, indicating that the SGP4 model does a fair job of modeling the satellite altitude. In the cross-track direction the variance grows slightly but in the transverse direction there is an order of magnitude larger growth in the error. As was noted in a majority of the references to the Literature Review, the axis of greatest error is expected to be in the transverse direction. In this direction the velocity and therefore drag components are greatest, so any modeling inaccuracies are amplified along this axis.

Finally, in the radial direction there appears to be a set of outliers for the whole period well above the variance of the mean. These may correspond to "bad" TLEs that had a large variance from the truth to start with and should be investigated further.

### 3.3.2 Error Topography

By plotting the range (the straight line distance) between every TLEs and truth on the same figure we can investigate the comparative growth of errors. For each propagated TLE in the period, the residual is plotted against the truth data at the epoch of every TLE in the period. Figure 3-3 shows the topography of the errors. Along the diagonal each TLE is propagated to its own time and compared against truth. This therefore gives the variance of the TLEs around the truth data.

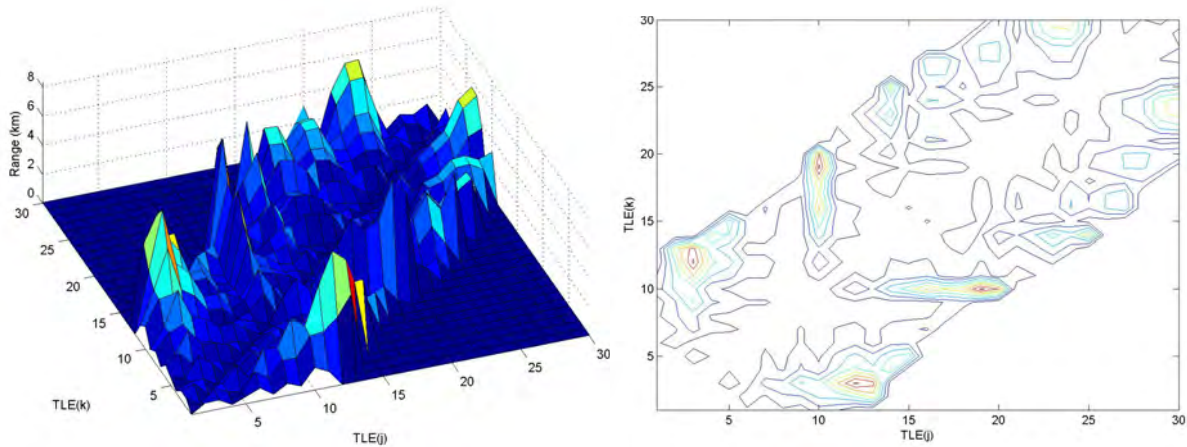

Figure 3-3: Stella range residuals for TLE(j) propagated forward in time to TLE(k) (limited to 7 days of propagation)

These figures give an idea of how certain TLEs diverge more rapidly from the truth data – it is these TLEs that are driving the spread of the variance in Figure 3-2 and identifying and excluding these may be one way of improving assessment of the TLE/propagator accuracies.

### 3.3.3 Modeling Statistical Quantities

The binning of the residuals discussed in Section 2.4 allows the bias and the error growth to be investigated further. Figure 3-4 shows the mean of each data bin and it is apparent that the mean in the radial and cross-track directions can be modeled fairly accurately by least squares fitting of a polynomial to the data. This allows a correction to be made of the residuals to remove the bias. The along-track direction does not allow for as good a fit however, and this should be repeated once bad TLEs have been identified and removed from the data to see if this gives a more predictable bias in the error. These biases are arising due to the inaccuracy of the SGP4 model and the exact source of inaccuracy should be investigated to identify SGP4's weakness.

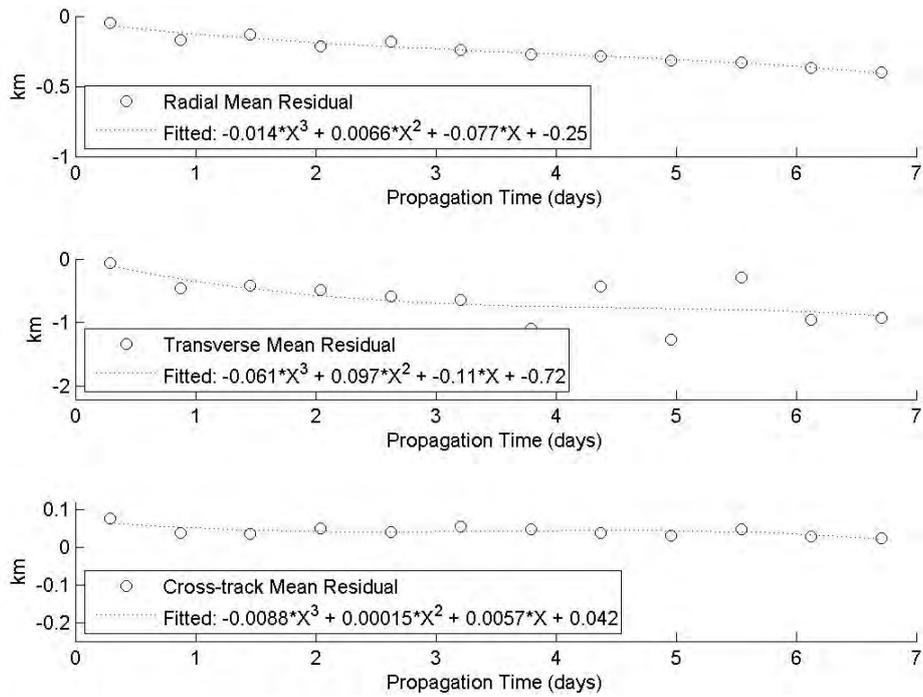

Figure 3-4: Mean residuals plotted for each data bin, with a least squares fitted 3${}^{rd}$ order polynomial

Repeating the same procedure for the standard deviation (or variance) gives an idea of how the error grows according to the SGP4 propagation time. Each of the plots in Figure 3-5 reflect the conclusions of the residual plots above – the radial and cross-track directions are roughly constant over 7 days of propagation but the transverse direction experiences a rapid growth in the error. Important to note is that by the end of the 7 days, the variance in the along-track direction is almost 16km. It is clear that after just a few days of propagation with SGP4 the errors are too big to allow informed maneuver decision making.

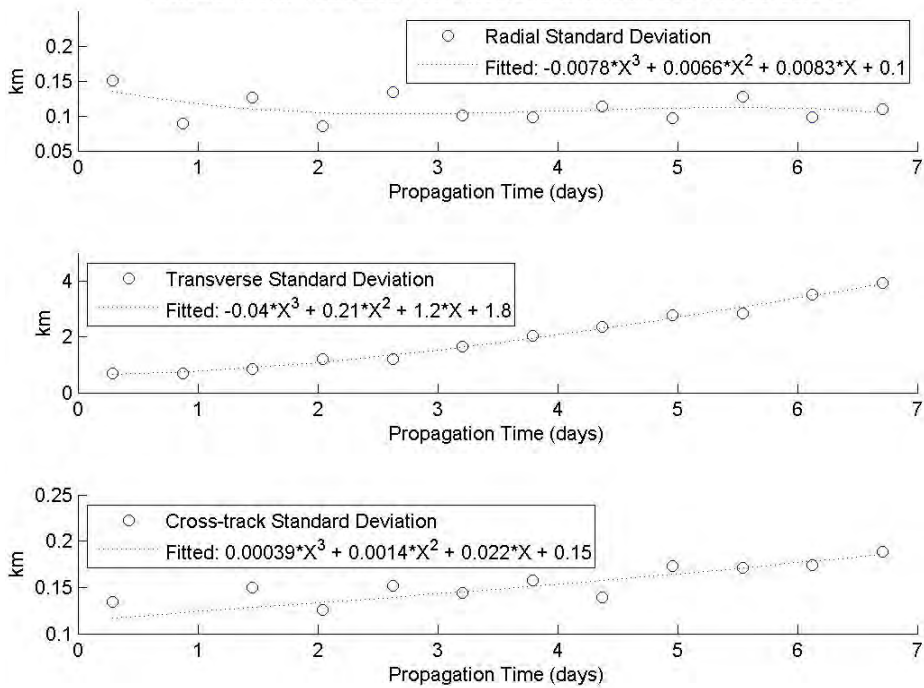

Figure 3-5: Standard deviations of residuals plotted for each data bin, with a least squares fitted 3${}^{rd}$ order polynomial

From these RTC errors an error ellipsoid can be generated under certain assumptions of the distribution of the data points. This error ellipsoid can then be propagated according to the polynomial fits of this historical data. This allows a way to propagate error ellipsoids that does not require the calculation of the state transition matrix of Section 2.5. While this is not totally independent of the force model selected because it is modeled on data from a chosen propagator it does provide an alternative and computationally less intensive way of propagating error.

This conversion from RTC probabilities to error ellipsoids was briefly investigated. Simply using the RTC standard deviations gives a conjunction rectangle and does not translate conjunction probabilities onto the standard error ellipsoid. An ellipsoid with equatorial radii *a, b* and polar radius *c* is defined in Cartesian coordinates *XYZ* by

$$\frac{x^2}{a^2} + \frac{y^2}{b^2} + \frac{z^2}{c^2} = 1 \tag{3-1}$$

For a one standard deviation (one-sigma, $\sigma$) scaled ellipsoid in RTC frame, with the center defined by the mean position, $\mu$, the inclusion of an actual residual *X* is tested with the condition:

$$\frac{(X_R - \mu_R)^2}{\sigma_R^2} + \frac{(X_T - \mu_T)^2}{\sigma_T^2} + \frac{(X_C - \mu_C)^2}{\sigma_C^2} < 1 \tag{3-2}$$

By applying this equation to each of the data bins and then averaging over the 12 bins the overall distribution of points was tested for a one sigma scaled ellipsoid. To have the standard one-sigma confidence level we require 67% of points to fall inside the ellipse. However, it was found that only 20% of residuals fell inside the ellipse at one-sigma. Scaling the ellipsoid by a factor of 1.83 gave a 67% confidence interval. A factor of 2.55 gave 95% confidence and a factor of 3.6 gave a 99.7 confidence interval – the requirements for a three-sigma validation of Gaussian distribution. This confirms that the distribution is non-Gaussian and this should be a topic for further investigation.

### 3.3.4 Evolution of Probability Density Function

To visualize the effect of the error growth and biases, Figure 3-6 shows the normal distribution density of the residuals against time. This is under the apparently erroneous assumption that the distribution is Gaussian. Inspection of the residuals in Figure 3-1 and the three-sigma analysis above confirm that the distribution is not normal, but making this assumption allows an easily understood visual representation of the situation to be generated.

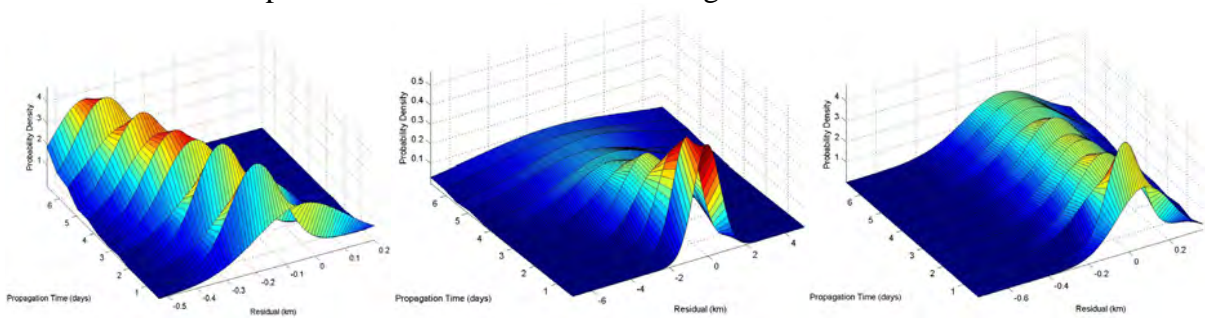

Figure 3-6: RTC Normal probability distribution functions, plotted for 7 days.
From left to right: Radial, Transverse and Cross-track

In the radial direction it is clear that the mean of the residuals drifts away from zero with the aforementioned bias, however the variance appears roughly constant. The same can be said for the cross-track direction, besides the slight peaking of the density for short propagations. In the along-track direction the rapid growth of the errors manifests as a spreading of the density function. This gives a useful impression of how, when propagating with SGP4, the uncertainty in satellite position becomes too large to manage after just a few days.

## 3.4  Summary

For objects with available truth data it is therefore possible to use this tool to investigate the uncertainty in the TLEs themselves and the uncertainty arising from propagating the TLEs into the future for conjunction analysis. The accuracy of the TLEs for Jan 2004 was of the order of 600m, with a standard deviation of about 340m, and this initial error appears to have a significant effect on the accuracy of the propagator. The major component of this error, like the velocity, is found in the along-track direction. The SGP4 propagator model is only useful for a few days of propagation before the variance is just too large to allow the uncertainty to be managed in the conjunction analysis decision space. The binning of residuals allows the bias and the error growth to be modeled and this may lead to ways to reduce the uncertainty. Finally, the distribution of the residuals appears to be non-Gaussian and should be investigated further.

# 4 Propagated TLEs Against TLEs in the Absence of Truth Data

## 4.1 Introduction to Task 0

In the case of debris objects no truth data is available for comparison. So in this case the variance cannot be accurately computed around the actual POEs. For this case the TLE at its epoch is defined as the best estimate of the satellite's position. It was determined previously that on average this is 600m from the actual position, but it is the best available estimate. The pair-wise differencing comparison can then be repeated under this assumption in the same fashion as before. This TLE vs. TLE investigation is routinely done by a number of systems (SOCRATES, MAESTRO and COVGEN) but is used in this analysis as the baseline for comparison as well as to provide Ames with its own in-house ability to do this comparison.

## 4.2 Results

These results are for Stella for the period of Jan 2004, where propagated TLEs are compared against TLEs at their own epoch.

### 4.2.1 Residuals

The residuals were calculated between the propagated TLEs and the unpropagated TLEs at each TLE epoch. Figure 4-1 plots out these residuals in the RTC frame. The residuals have similar general characteristics as those from Figure 3-2 – there is a bias in the radial direction, a rapidly growing variance in the along-track direction and a slowly growing variance in the cross-track. The magnitude of these errors is about the same as for the previous comparison against truth. The effect of the method of comparing TLEs to themselves is that the variance at time zero, with no propagation, is zero – the variance of the actual error in the TLEs has disappeared.

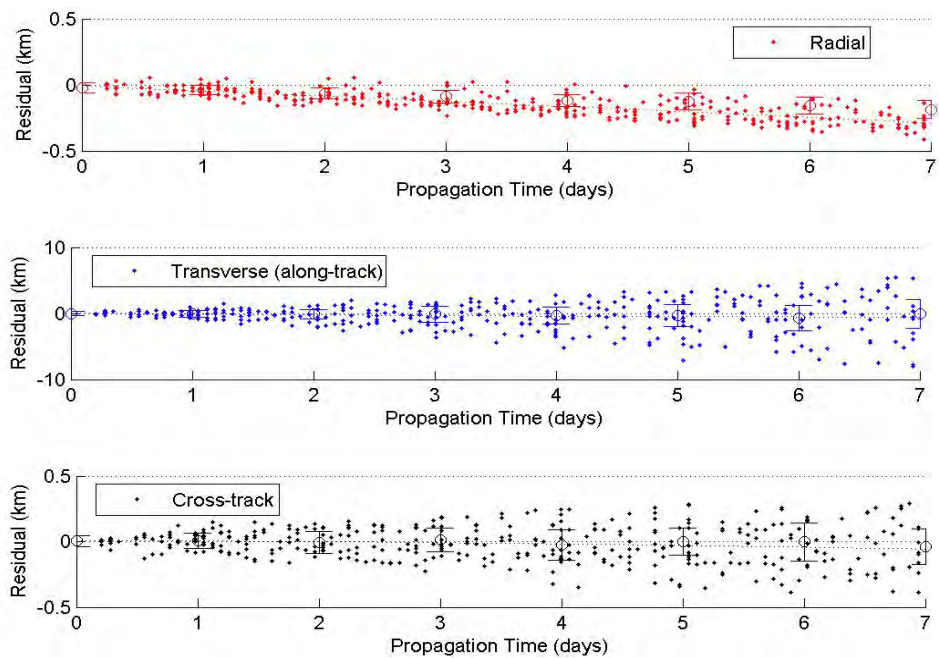

Figure 4-1: Stella RTC position residuals for Jan 2004 TLEs propagated to TLEs for 7 days of propagation

## 4.2.2 Error Topography

Analyzing the error topography resulting from propagating every TLE to every other TLE's epoch and comparing the residuals actually gives a clearer image of error growth than comparing to truth. This is because the variability in the TLEs around truth is removed, effectively decluttering the topography.

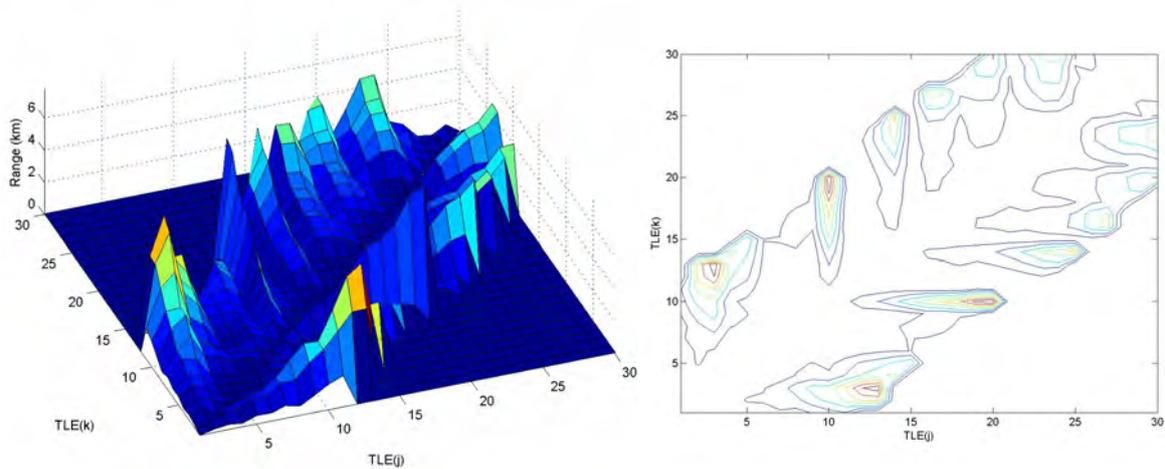

Figure 4-2: Stella range residuals for TLE(j) propagated forward in time to TLE(k) (limited to 7 days of propagation)

The result, show in Figure 4-2, is more representative of error growth directly resulting from different initial conditions in the SGP4 propagator. The diagonal in this case is zero – the TLE propagated to itself – but the divergent propagation paths can be identified even more clearly. A full analysis of which TLEs these are and why they diverge faster was not conducted, but this would be the approach of a "bad TLE" identification algorithm.

## 4.2.3 Modeling Statistical Quantities

As previously, the mean biases and error growth can be modeled by a least squares polynomial fit, shown in Figure 4-3. There appears to be a near-linear radial bias driven by the SGP4 propagator for the Stella satellite. The along-track and cross-track directions are less easy to model, but still display some predictable pattern.

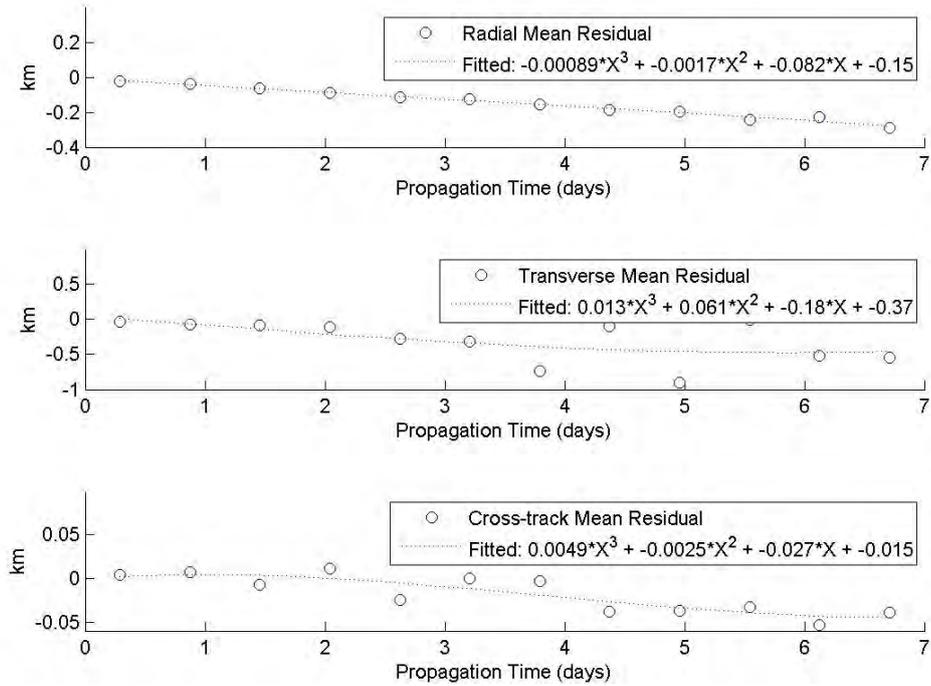

Figure 4-3: Mean residuals plotted for each data bin,
with least squares fitted 3$^{rd}$ order polynomial

The best estimate error growth, determined from calculating residuals between SGP4 propagated TLEs and the TLEs themselves, reveals similar trends to the previous case with the primary difference being that the variance logically approaches zero as time tends to zero. Consulting Figure 4-4 shows that this is not clearly the case in the radial direction, where the variance stays roughly constant and small for the period.

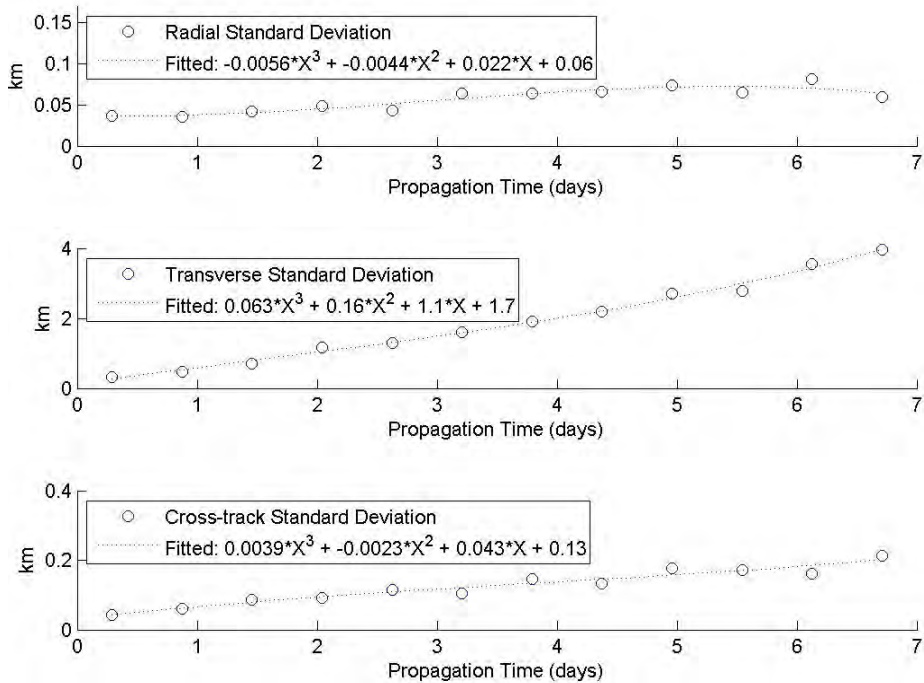

Figure 4-4: Standard deviation of residuals plotted for each data bin,
with least squares fitted 3$^{rd}$ order polynomial

It was found that only 17% of residuals fell inside the ellipse at one-sigma. Scaling the ellipsoid by a factor of 1.76 gave a 67% confidence interval. A factor of 2.7 gave 95% confidence and a factor of 3.7 gave a 99.7 confidence interval – the requirements for a three-sigma validation of Gaussian distribution. As before this suggests a non-Gaussian distribution which should be a topic for further investigation, along with how to determine this scaling factor.

### 4.2.4 Evolution of Probability Density Function

Once again, investigation of the evolution of the assumed normal probability distribution function reveals the dispersion of uncertainty in the along-track direction and the clear bias in the radial direction. The usefulness of this assumed distribution function is debatable, but presented in Figure 4-5 nonetheless.

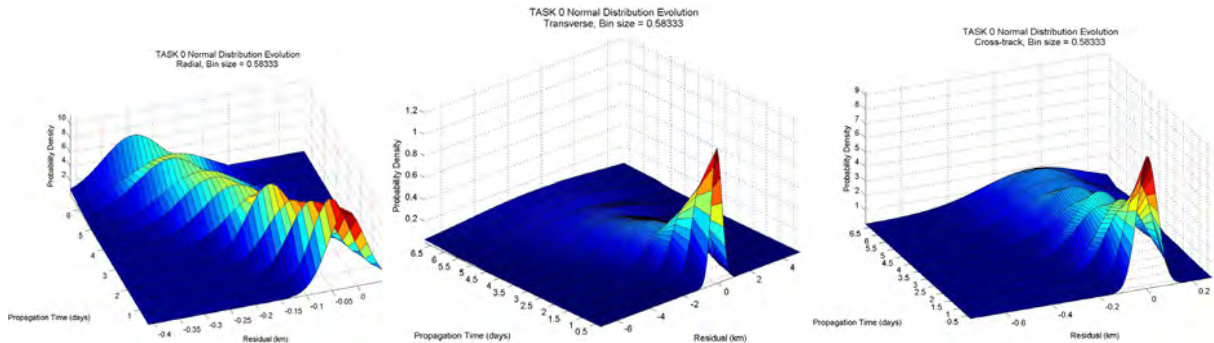

Figure 4-5: RTC Normal probability distribution functions, plotted for 7 days

### 4.2.5 Comparison Between TLEs as a Best Estimate and Known Truth Data

In order to compare the actual errors, determined by calculating residuals between SGP4 propagated TLEs and truth data, with the best estimate errors discussed here the variance of each are plotted in Figure 4-5.

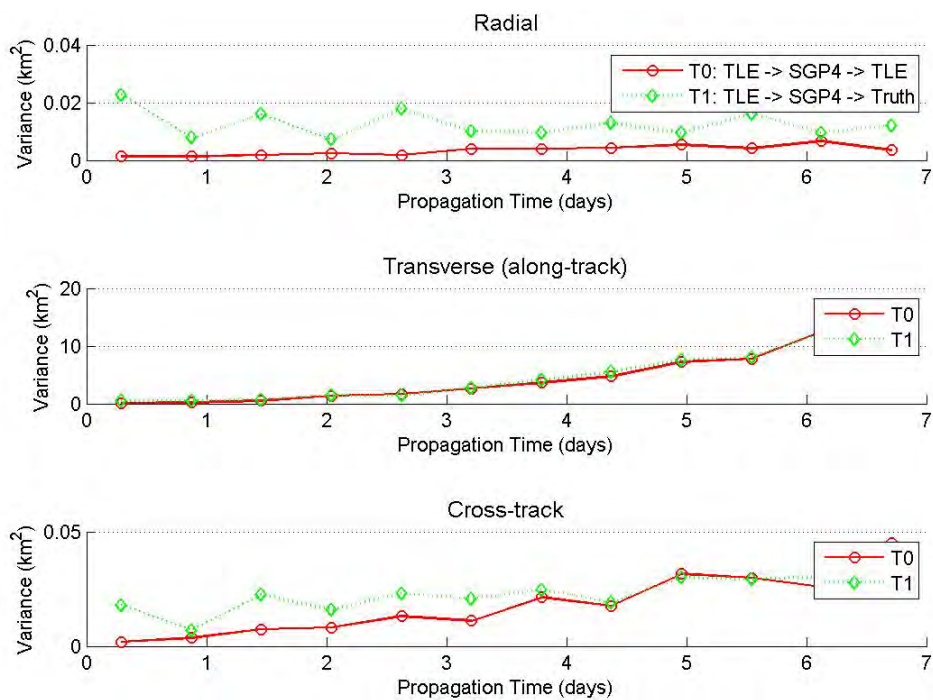

Figure 4-6: RTC variances for Task 0 and Task 1

There are several points to note from this figure. The sizes of the variances are very similar, regardless of whether you know truth data or use the TLE as a best estimate. In the radial direction there appears to be a constantly higher variance when comparing against truth and this could be a result of the outliers discussed with Figure 3-2. Removal of these outliers may result in very similar variances. In the along-track direction, at least at this scale, the variances are indistinguishable. In the cross-track direction the variance against truth starts out greater, but approaches that of the TLE best estimate after about 4 days.

This suggests, for the along-track and cross track directions, that when propagating for more than 4 days it is irrelevant, in terms of errors, whether or not truth data is known. This can only be applied to Stella for this period at the moment, but should be investigated further. Separately, correction of bad TLEs may lead to a similar conclusion for the radial direction.

## 4.3 Summary

For objects, such as debris, where no truth data is available it is necessary to use the unpropagated TLEs as best estimates of satellite position. The residuals display similar characteristics regardless of whether truth data is known, suggesting that the propagator is the source of the bias found in the radial direction. TLEs as a best estimate is the method employed elsewhere and an analysis of the errors shows that using this assumption does not significantly underestimate the actual uncertainties. With no propagation, the difference is the variance in the TLEs around truth, but after just a few days of propagating this is swamped by the variance introduced by the SGP4 propagator.

# 5 HPOP as an Alternative Propagator

## 5.1 Introduction to Task 2

This analysis is extended to investigate STK's High Precision Orbit Propagator (HPOP). HPOP works by numerically integrating the equations of motion to generate ephemeris. The propagator allows various different force modeling effects to be included as well as using different numerical integration techniques. Different parameter settings make HPOP the most accurate STK propagator, provided the user knows and enters the appropriate initial conditions and force model settings (Vallado, 2005). For this reason HPOP is not strictly appropriate for propagating TLEs – the TLEs are mean orbital elements and are not accurate enough to use as initial conditions for HPOP. Additionally, for orbital debris the properties of the object are not known sufficiently to accurately model the forces involved. However, it is useful to quantify this inappropriateness as well as to demonstrate the use of an alternative propagator in the MATLAB/STK tool. Each TLE was converted into a state vector, which was used to initialize the HPOP propagator. These were propagated to the end time and residuals compared against the unpropagated state vectors in a pair-wise fashion.

For the Stella spacecraft the area to mass ratio, the solar radiation pressure coefficient, the drag coefficient and the satellite mass were explicitly defined in the force models. The Jacchia-Roberts atmospheric model was used which computes atmospheric density based on the composition of the atmosphere and includes some analytical enhancements to improve performance. This model depends on the satellite's altitude as well as including and seasonal variations. Numerical integration is done using $7^{th}$-$8^{th}$ order Runge-Kutta-Fehlberg which allows for good accuracy, but results in increased computational requirements for the HPOP model.

## 5.2 Results

Figure 5-1 is an STK screenshot, showing the variation in the positions of the propagated satellites after 30 days. The (yellow) HPOP satellites are significantly more spread out than the (blue) SGP4 satellites, suggesting that the initial conditions or parameters are amplifying the difference between these models, at least in the along-track direction.

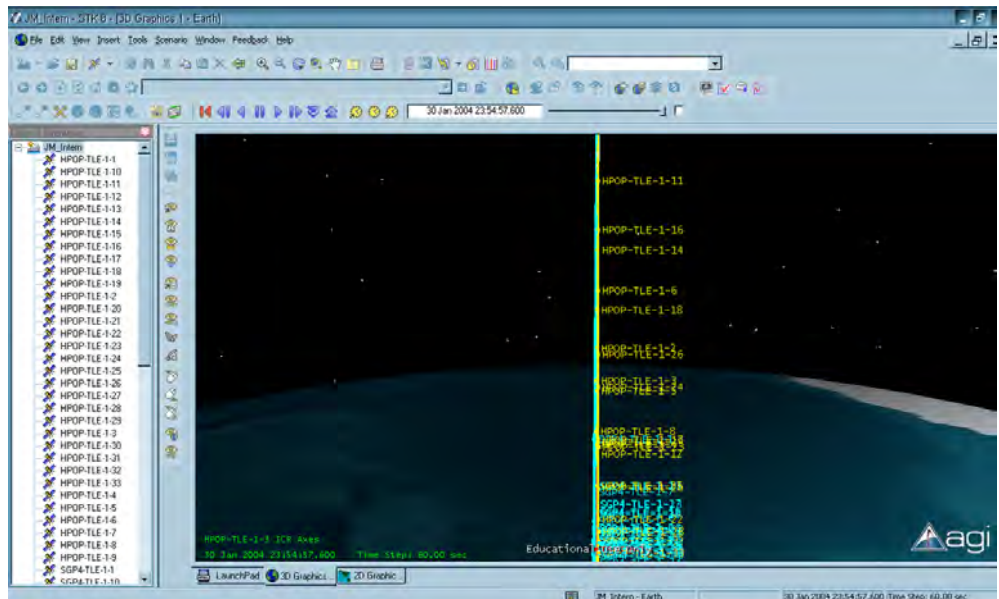

Figure 5-1: STK screenshot showing spread of HPOP propagated TLEs

### 5.2.1 Residuals

The RTC residuals shown in Figure 5-2 (with transverse at the top) show that the along-track residuals are an order of magnitude larger than those produced with the SGP4 propagator. However, in the radial and cross-track directions the residuals are the same order of magnitude, indicating that the HPOP model is most sensitive in the along-track direction. Initially the drag parameters were thought to be the source of this huge dispersion and the code was rerun with the drag parameter set to 0.1, 1, 2 and 2.2, but the reduction in the observed dispersion in the along-track position was found to be less than 2% regardless of the parameters. Additionally the atmospheric model was changed to the CIRA 1972[1] model with a similarly small effect on propagated orbit.

---

[1] CIRA is an empirical model of atmospheric temperature and densities as recommended by the Committee on Space Research (COSPAR) and formalized in 1972.

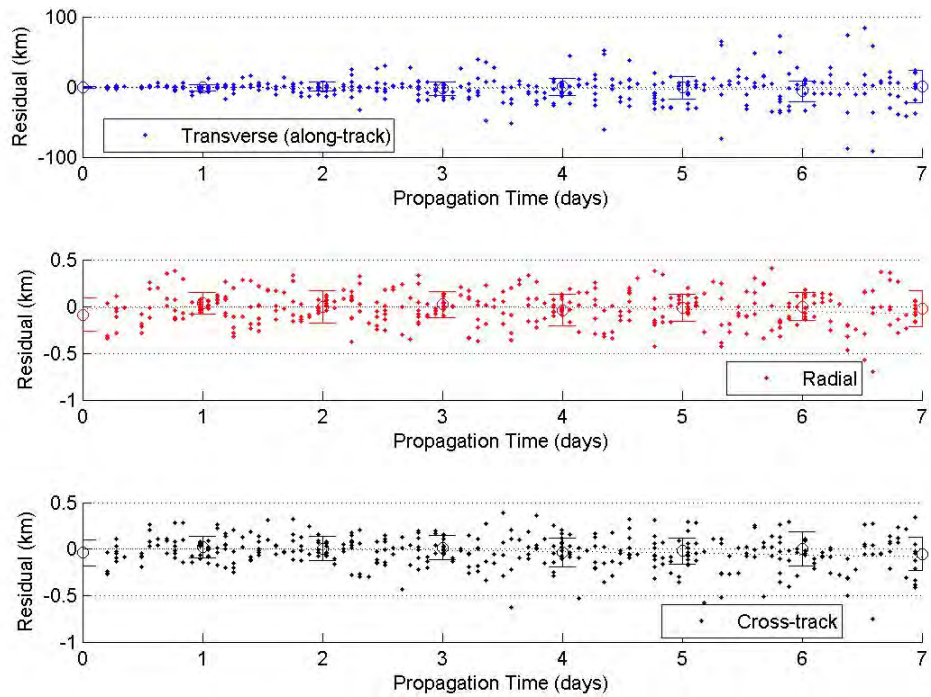

Figure 5-2: Stella RTC position residuals between HPOP propagated TLEs for 7 days of propagation

### 5.2.2 Error Topography

The error topography in Figure 5-3 can be plotted as before, with the advantage that HPOP has amplified all of the range residuals and given the topography a more defined relief. This may lend itself even more effectively to identifying the divergent TLEs. However, more work needs to be done in order fully understand this and to apply it effectively.

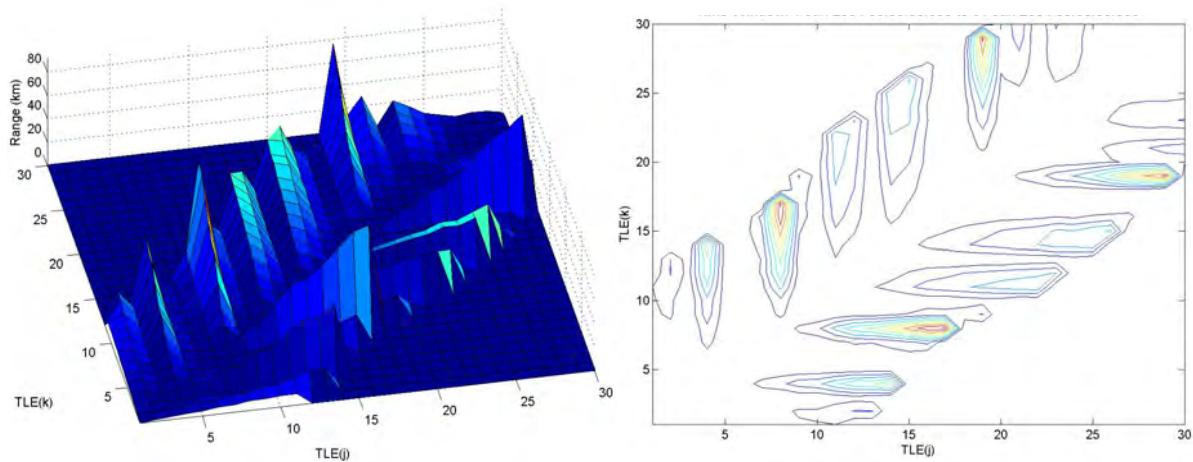

Figure 5-3: Stella range residuals for TLE(j) propagated forward in time to TLE(k) (limited to 7 days of propagation)

### 5.2.3 Comparison to SGP4 Methods

To further assess HPOP in relation to SGP4, the variances are plotted in RTC in Figure 5-4. The variance is slightly lager in the cross-track direction and does not approach the comparison against truth as T1 did. In the radial direction, the variance is also slightly larger, but in the along-track direction the error grows hugely in comparison to the SGP4 propagator errors.

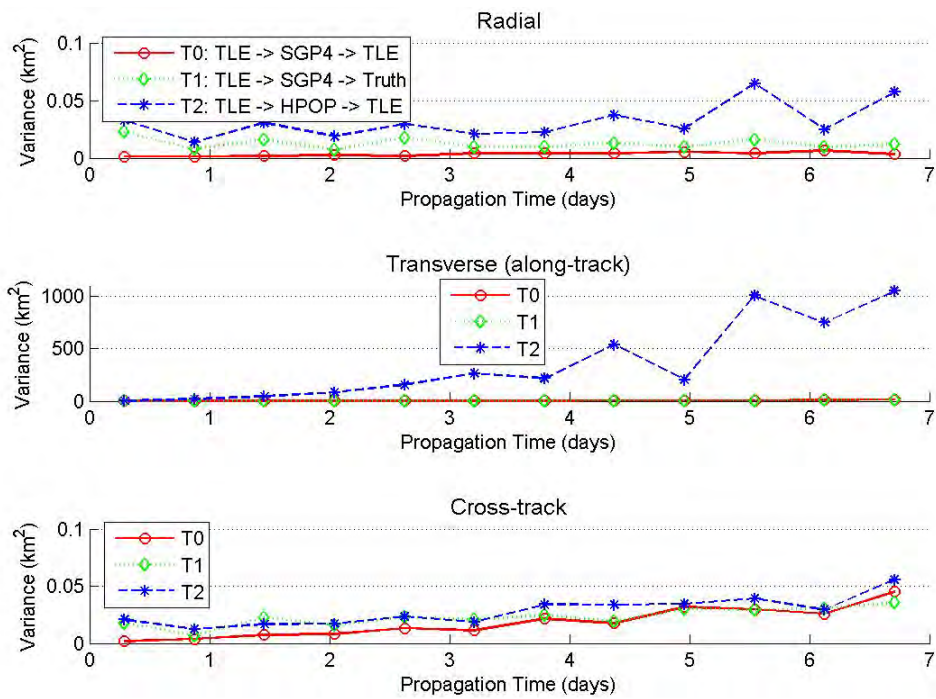

Figure 5-4: RTC variances for Task 0, Task 1 and Task 3

## 5.3 Summary

It appears as if the variability in the TLEs when they are converted to state vectors is sufficient to cause variances of the order of thousands of kilometers in just a week. The variance caused by parametrically altering the force model parameters does have an effect, but not nearl as large. This simply confirms that the HPOP propagator's sensitivity to initial state vector conditions renders it inappropriate for use with TLEs. Additionally it demonstrates that a different propagator can be assessed using this pair-wise differencing machinery and compared against the SGP4 standard to quantify their relative performance at propagating satellites and their associated positional uncertainties.

# 6  Conclusions

The purpose of this whole endeavor was to develop a rapid prototyping and testing environment for comparing different strategies for improving the accuracy of conjunction analysis using the public TLE catalog. The scripting of the MATLAB/STK interface allows an assessment of the accuracy of the TLEs compared to actual truth data, the TLEs compared to a best estimate (the TLEs themselves) and of the propagator chosen to predict into the future. The current implementation takes in two data sources and uses two STK-integrated propagators to compare their relative accuracies using the pair-wise differencing method.

The system allows new data inputs to be simply integrated. For example, a set of TLEs that has been corrected by some means could be fed into the system and the results analyzed to determine the magnitude of the improvement resulting from the correction. Similarly, additional STK propagators can be integrated easily into the code and comparisons conducted of their uncertainty propagation.

## 6.1  Summary of Stella Results

A case study to demonstrate the system's capabilities was conducted for the CNES Stella satellite by comparing TLEs against truth and investigating how the errors grow with propagation time. Additionally the covariance matrices were generated for each case. Although these are not discussed in detail they are presented in Appendix C and discussed further in the literature (Osweiler, 2006).

It was found that the TLEs for Jan 2004 are on average 600m away from the actual satellite positions. When propagating TLEs into the future using SGP4 or HPOP the largest errors appear in the along-track direction. The growth of these errors appears to have some order and can be modeled and possibly projected into the future in the form of an error ellipsoid. Similarly, biases in the residuals can be modeled and removed from the data, which can improve position predictions. After propagating for about 4 days, the errors of the TLEs around themselves are roughly the same magnitude as the TLEs around truth, suggesting that the use of TLEs as a best estimate of truth is a valid assumption for Stella after 4 days of propagation.

However, even if truth data is available for an object, using the uncorrected TLEs with the SGP4 propagator results in variances in position of the order of tens of kilometers after a week. This is too much uncertainty to allow for decision making and it is clear that further strategies need to be developed to improve the accuracy of the TLE/SGP4 system.

## 6.2  Impact on project

The machinery developed here, namely the implementation of pair-wise differencing using multiple data sources and propagators, allows proposed strategies to be tested before they are implemented on a larger scale. In this environment externally corrected or enhanced TLE data can be used as an input to determine the magnitude of any improvement. Similarly, various propagators can be tested – STK integrated propagators can be plugged in seamlessly and new propagators can be added to the code as necessary. Once effective strategies have been identified they can be streamlined and redeveloped for implementation on the Pleiades supercomputer across the whole catalog.

Additionally, the covariance matrices of a particular satellite can be determined using this method, either in relation to truth data or just from the TLEs. This covariance data can be used with a state transition matrix to propagate error ellipsoids on the supercomputer's conjunction analysis simulation. At the moment, the simulation uses miss distance "hard spheres" and replacing these with actually calculated error ellipsoids would allow collision risks to be quantified. Alternatively, the modeling of error growth presented here is a way to generate error ellipsoids that depends on the propagator used to do pair-wise differencing but does not depend on a calculation of the future state matrix.

Ultimately the purpose of this tool is a development and exploration environment to support the assessment of strategies to be implemented in a decision tool based on the Pleiades supercomputer.

## 6.3 Future Work

A number of opportunities to enhance and improve this code were suggested as well as some avenues for further investigation. These are broken down into 'expansion of the tool' and 'future research' and are presented below.

### 6.3.1 Expansion of the Tool

The addition of new propagators would allow further strategies to be investigated. This implementation allows for easy addition of new STK-integrated propagators, but the code should be expanded to allow for additional plug-in propagators to be evaluated.

Similarly, new data sources should be identified and integrated into the system. The most likely example would be corrected TLEs, that have had some of their variance removed through smoothing or fitting of historical data. This would then allow the resulting reduction in propagated error to be analyzed.

TLEs that result in highly divergent residuals need to be identified and removed from the historical data. This could be done by backwards comparison of TLE to identify statistical outliers as is done by GOODOB (Kaya, 1999). An alternative way would be to further investigate the range topography plots presented in this report and developing a criteria for TLE removal based on the range growth rates.

The modeling of the bias in the radial residuals for Stella provides a method of first order corrections by removing this bias from the data. While this does not improve the variance of the dataset, it does move the mean of the data and ensure that the error ellipsoids are centered on the actual mean and not a propagator biased psuedomean.

### 6.3.2 Future Research

The code should be expanded to generate error ellipsoids from the covariance matrix or from the modeled RTC errors. This is a natural addition that would be beneficial before implementation on the supercomputer. The similarities between the two methods of ellipsoid generation and an investigation of the underlying assumptions might provide insight into which method would prove better for implementation across the whole catalog.

The actual distribution of the residual data should be further investigated because it is clear that the assumption of a normal distribution is not strictly valid. An accurate assessment of the true distribution is critical for determining a valid probability distribution function and for generating realistic error ellipsoids.

Finally, the investigation of different accuracy improving strategies and their effective implementation on the supercomputer is the key area of future work and this tool should assist in comparing and evaluating competing approaches.

# 7 List of References

# 8 Appendix

## Appendix A                      TLE Format Description

This description is taken from CelesTrak.com:

Data for each satellite consists of three lines in the following format:

```
AAAAAAAAAAAAAAAAAAAAAAAA
1 NNNNNU NNNNNAAA NNNNN.NNNNNNNN +.NNNNNNNN +NNNNN-N +NNNNN-N N NNNNN
2 NNNNN NNN.NNNN NNN.NNNN NNNNNNN NNN.NNNN NNN.NNNN NN.NNNNNNNNNNNNNN
```

Line 0 is a twenty-four character name (to be consistent with the name length in the NORAD SATCAT).

Lines 1 and 2 are the standard Two-Line Orbital Element Set Format identical to that used by NORAD and NASA. The format description is:

**Line 1**

| Column | Description |
|---|---|
| 01 | Line Number of Element Data |
| 03-07 | Satellite Number |
| 08 | Classification (U=Unclassified) |
| 10-11 | International Designator (Last two digits of launch year) |
| 12-14 | International Designator (Launch number of the year) |
| 15-17 | International Designator (Piece of the launch) |
| 19-20 | Epoch Year (Last two digits of year) |
| 21-32 | Epoch (Day of the year and fractional portion of the day) |
| 34-43 | First Time Derivative of the Mean Motion |
| 45-52 | Second Time Derivative of Mean Motion (decimal point assumed) |
| 54-61 | BSTAR drag term (decimal point assumed) |
| 63 | Ephemeris type |
| 65-68 | Element number |
| 69 | Checksum (Modulo 10) (Letters, blanks, periods, plus signs = 0; minus signs = 1) |

**Line 2**

| Column | Description |
|---|---|
| 01 | Line Number of Element Data |
| 03-07 | Satellite Number |
| 09-16 | Inclination [Degrees] |
| 18-25 | Right Ascension of the Ascending Node [Degrees] |
| 27-33 | Eccentricity (decimal point assumed) |
| 35-42 | Argument of Perigee [Degrees] |
| 44-51 | Mean Anomaly [Degrees] |
| 53-63 | Mean Motion [Revs per day] |
| 64-68 | Revolution number at epoch [Revs] |
| 69 | Checksum (Modulo 10) |

All other columns are blank or fixed.

Example:

```
NOAA 14
1 23455U 94089A   97320.90946019  .00000140  00000-0  10191-3 0  2621
2 23455  99.0090 272.6745 0008546 223.1686 136.8816 14.11711747148495
```

**Appendix B          RTC Coordinate Frame**

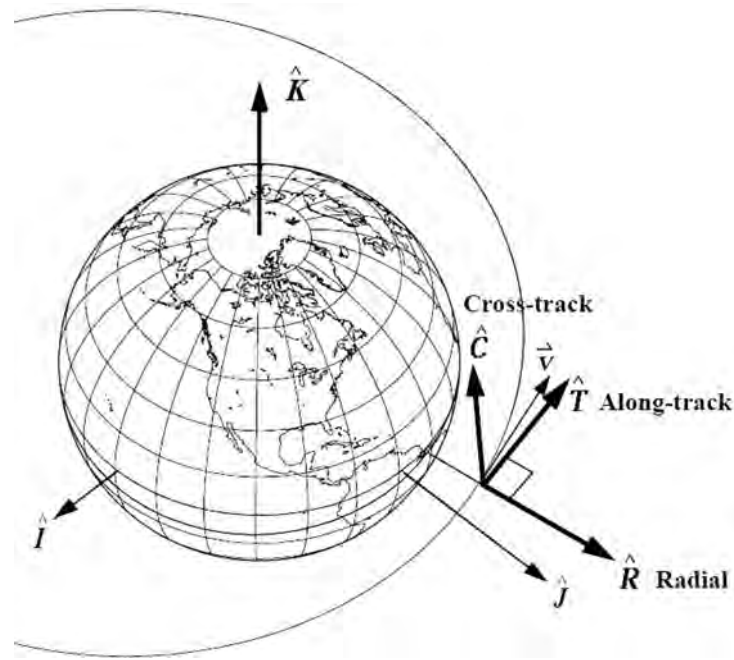

Figure 8-1: RTC frame definition (Vallado, 2003)

Definition adapted from Osweiler, 2006:

The RTC system (for radial, transverse, and cross-track) is a natural satellite-based coordinate system – it moves with the satellite. It is also known as RSW or RTN. The R axis always points from the Earth's center along the radius vector (position vector) toward the satellite, as it moves through its orbit. Dividing by its magnitude provides the R unit vector. The C axis is found by crossing the radius vector into the velocity vector, and normalizing it gives the C unit vector. Finally, the direction of the T axis and the T unit vector are found by crossing the C and R unit vectors. The T axis is in the orbital plane, perpendicular to the radius vector.

The transverse axis (T) points in the direction of, but not necessarily parallel to, the velocity vector. It is exactly collinear to the velocity vector for circular orbits or at apogee and perigee for elliptical orbits. Also, differences measured in the transverse direction are commonly known as along-track displacements, or errors, and differences measured in the C axis direction are known as cross-track errors.

# Appendix C            Covariance Matrices

The calculated covariance matrices for Stella for January 2004, with propagations limited to 7 days. The covariances given here are in classical orbital elements, with:

*a* – semi major axis
*i* – inclination
*e* – eccentricity
*Ω* - longitude of the ascending node
*ω* - argument of perigee
*M* – mean anomaly

**Task 0:** Propagating TLEs with SGP4 and comparing them to TLEs as a best estimate:

|   | *a* | *i* | *e* | *Ω* | *ω* | *M* |
|---|---|---|---|---|---|---|
| *a* | 0.006026 | -8.7E-09 | -1.8E-09 | 1.12E-08 | 5.65E-06 | -5.9E-06 |
| *i* | -8.7E-09 | 3.2E-14 | 1.34E-14 | -3.6E-14 | -8.7E-12 | 8.76E-12 |
| *e* | -1.8E-09 | 1.34E-14 | 1.48E-14 | 2.93E-15 | 1.78E-12 | -1.6E-12 |
| *Ω* | 1.12E-08 | -3.6E-14 | 2.93E-15 | 7.96E-14 | 1.85E-11 | -1.8E-11 |
| *ω* | 5.65E-06 | -8.7E-12 | 1.78E-12 | 1.85E-11 | 7.72E-09 | -7.9E-09 |
| *M* | -5.9E-06 | 8.76E-12 | -1.6E-12 | -1.8E-11 | -7.9E-09 | 8.13E-09 |

Task 1: Propagating TLEs with SGP4 and comparing them known truth data:

|   | *a* | *i* | *e* | *Ω* | *ω* | *M* |
|---|---|---|---|---|---|---|
| *a* | 0.003308 | 1.01E-08 | -9.1E-09 | -8.3E-09 | -3.5E-06 | 3.3E-06 |
| *i* | 1.01E-08 | 2.49E-13 | -2.3E-13 | -2E-13 | -1.5E-10 | 1.46E-10 |
| *e* | -9.1E-09 | -2.3E-13 | 2.43E-13 | 2.16E-13 | 1.48E-10 | -1.5E-10 |
| *Ω* | -8.3E-09 | -2E-13 | 2.16E-13 | 2E-13 | 1.29E-10 | -1.3E-10 |
| *ω* | -3.5E-06 | -1.5E-10 | 1.48E-10 | 1.29E-10 | 9.3E-08 | -9.3E-08 |
| *M* | 3.3E-06 | 1.46E-10 | -1.5E-10 | -1.3E-10 | -9.3E-08 | 9.31E-08 |

Task 2: Propagating TLEs wth HPOP and comparing them to TLEs as a best estimate:

|   | *a* | *i* | *e* | *Ω* | *ω* | *M* |
|---|---|---|---|---|---|---|
| *a* | 0.174671 | 2.19E-08 | -1.5E-07 | 1.36E-08 | -5.7E-05 | 5.46E-05 |
| *i* | 2.19E-08 | 1.7E-14 | 4.28E-16 | 9.29E-15 | -1.2E-11 | 1.15E-11 |
| *e* | -1.5E-07 | 4.28E-16 | 2.02E-13 | 4.31E-14 | 3.45E-11 | -3.5E-11 |
| *Ω* | 1.36E-08 | 9.29E-15 | 4.31E-14 | 8.23E-14 | -2.6E-11 | 2.35E-11 |
| *ω* | -5.7E-05 | -1.2E-11 | 3.45E-11 | -2.6E-11 | 3.21E-08 | -3.1E-08 |
| *M* | 5.46E-05 | 1.15E-11 | -3.5E-11 | 2.35E-11 | -3.1E-08 | 3.03E-08 |